\documentclass[prl,aps,twocolumn,showpacs]{revtex4}
\usepackage{graphicx}
\usepackage{dcolumn}

\begin{document}

\title{Vortex Core Structure in Neutral Fermion Superfluids with Population Imbalance}

\author{M. Takahashi}
\affiliation{Department of Physics, Okayama University,
Okayama 700-8530, Japan}
\author{T. Mizushima}
\affiliation{Department of Physics, Okayama University,
Okayama 700-8530, Japan}
\author{M. Ichioka}
\affiliation{Department of Physics, Okayama University,
Okayama 700-8530, Japan}
\author{K. Machida}
\affiliation{Department of Physics, Okayama University,
Okayama 700-8530, Japan}
\date{\today}

\begin{abstract}

Quantized vortex core structure is theoretically 
investigated in Fermion
superfluids with population imbalance for two atom species
of neutral atom clouds near a Feshbach resonance.
In contrast with vortex core in balance case where
the quantum depletion makes a vortex visible through
the density profile measurement, the vortex core is filled in and becomes less visible because the quantized discrete 
bound states are occupied exclusively by the majority species. Yet it is shown that the core can be visible through the minority density profile experiment
using phase contrast imaging, revealing an interesting opportunity to examine low-lying Fermionic core bound states
unexplored so far.

\end{abstract}


\maketitle

There has been much attention focused on Fermionic superfluids of cold atoms, such as $^6$Li or $^{40}$K since experimental realization \cite{review}. The experiments utilize the Feshbach resonance by changing an external field to tune the atom-atom interaction, achieving the BEC-BCS crossover. 
A keen interest is now placed on Fermion superfluidity when the population of the two species (up and down ``spins'') is unequal experimentally \cite{mit1,mit2,mit3,rice} and theoretically \cite{machida06,kinnunen}. Zwierlein {\it et al.} \cite{mit1,mit2} have succeeded in creating vortices not only in the balance case with equal population, but also in the imbalance case, directly demonstrating its superfluidity. Here the presence or absence of vortices in a system is utilized to monitor superfluidity because quantized vortex is a hallmark of superfluidity. They have observed the clear signature of vortices in the inner region of the density after sweeping into the BEC side, even in the imbalance case of the BCS side. However, it is important to notice that whether the superfluidity in the outer region is robust or not is still open to question. In this Letter, we focus on the visibility of the vortex core situated at the trap center in the imbalance case.
In addition, we discuss the visibility of vortices in the outer region from the results. This result is quite contrasted in the BEC case where vortices are remarkably arranged regularly throughout whole system, even near the boundary with lower density \cite{vortices}.

The order parameter  is $\Psi _{\rm BEC} = \sqrt{n(r)}$ ($n(r)$ is the number density) in BEC, while the Cooper pair amplitude $\Psi _{\rm BCS} \propto \langle c^{\dag}_{k \uparrow}c^{\dag}_{-k \downarrow}\rangle$ in BCS. Thus it is not self-evident even in the $50\%$-$50\%$ balance case that the vortex, which is probed by the density contrast, continues to be ``seen'' from BEC to BCS  across the Feshbach resonance point. A theoretical question here is to understand the peculiar vortex core structure in the imbalance case, related to the quantum depletion of the density at the core \cite{feder,hayashi}.  We provide a microscopic calculation for it, fully taking account  of low-lying Fermionic excitations around a core beyond simple local density approximation. These individual excitations are essential in Fermion superfluid, a feature completely absent in the vortex state in Boson superfluid.

This study might be useful for other research fields such as condensed matter community because superconductors with imbalance up and down spin electrons are realized by an applied field through the Zeeman effect. Thus there is a good chance to check the present study. Indeed in a heavy Fermion material CeCoIn$_5$ the ``imbalance'' superconductivity may be realized under high fields where the Fulde-Ferrell-Larkin-Ovchinnikov (FFLO) state is observed \cite{kakuyanagi1}. 
Also in high temperature superconductors the so-called ``vortex charging'' problem where the quasi-particle density is suppressed at the core as observed by nuclear quadrupole resonance (NQR) experiment \cite{kakuyanagi2} and  anomalous Hall conductivity \cite{hall}.
We also expect that the study of the vortex core in the imbalance superfluid might be useful for the inside of rotating neutron stars where vortices are sustained in imbalance color superconductivity, consisting of quark matter \cite{RMP}.

Before going into detailed calculation in the imbalance case, we recapture the results in the balance case; The core structure  both in superconductors and superfluids
is studied in the balance case \cite{feder,hayashi}. 
They show that 
as increasing the coupling constant of attractive
interaction, the particle number at the core decreases due to quantum depletion. This is caused by the fact that the vortex core bound states, namely,
Caroli-de Gennes-Matricon (CdGM) states \cite{caroli} 
with finite amplitude at the vortex center are
unoccupied by discretization \cite{feder,hayashi}.
It should be noted, however, that there is no experiment
to directly detect the CdGM state in condensed matter physics so far. Here we show through this study that there is a good chance to directly see it.

To address such a problem, we self-consistently solve the Bogoliubov-de Gennes (BdG) equation  describing the interacting Fermion system with population imbalance under a trap $V({\bf r})$.  This framework is one of the most fundamental microscopic theories. We start with the BdG equation for the quasi-particle wave functions $u_{\bf q}({\bf r})$ and $v_{\bf q}({\bf r})$ labeled by the quantum number ${\bf q}$ as follows \cite{mizushima1,mizushima2}: 
\begin{eqnarray}
\left[ 
	\begin{array}{cc}
		\mathcal{K}_{\uparrow} (r) & \Delta (r) \\
		\Delta ^{\ast} (r) & - \mathcal{K}^{\ast}_{\downarrow}(r) 
	\end{array}
\right] 
\left[ 
	\begin{array}{c} u_{\bf q}({\bf r}) \\ v_{\bf q}({\bf r}) \end{array}
\right] =  E _{\bf q}
\left[ 
	\begin{array}{c} u_{\bf q}({\bf r}) \\ v_{\bf q}({\bf r}) \end{array}
\right],
\label{eq:bdg}
\end{eqnarray}
where $\mathcal{K}_{\uparrow,\downarrow} (r) = - \frac{\nabla^2_{\bf r}}{2M} +V(r) + gn _{\downarrow,\uparrow} (r) - \mu _{\uparrow,\downarrow}$. We fully takes account  of the mismatch Fermi surfaces as $\delta\mu \equiv (\mu _{\uparrow} - \mu _{\downarrow})/2$ and the local Hartree potential $g n_{\sigma}({\bf r})$ with $n_{\uparrow}({\bf r})= \sum _{\bf q} |u_{\bf q}({\bf r})|^2 f(E_{\bf q})$ and $n_{\downarrow}({\bf r})= \sum _{\bf q} |v_{\bf q}({\bf r})|^2 [1- f(E_{\bf q})]$. While eigenstates and eigenvalue of Eq. (1) are often written as $(u_{{\bf q}', \uparrow}, v_{{\bf q}', \uparrow})$, $E_{{\bf q}', \uparrow}$ and $(-v^\ast_{{\bf q}', \downarrow}, u^\ast_{{\bf q}', \downarrow})$, $-E_{{\bf q}', \downarrow}$, we express these states as $(u_{\bf q},v_{\bf q})$, $E_{\bf q}$. The bare attractive interaction $g = 4 \pi a /M$ via the scattering length $a$. Throughout this paper, we set $\hbar = 1$ and consider a cylindrical system with $V({\bf r})=\frac{1}{2}M\omega^2r^2$, imposing a periodic boundary condition with the periodicity $Z=3d$ ($d^{-1} = \sqrt{M\omega}$) along the $z$-direction. Hence, we write the wave functions as $u_{\bf q} ({\bf r})=u_{\bf q}(r)\exp{[i(\ell - \frac{1}{2})\theta + iq_z z]}$ and $v_{\bf q} ({\bf r})=v_{\bf q}(r)\exp{[i(\ell + \frac{1}{2})\theta + iq_z z]}$ with $q_z = 0, \pm 2\pi/Z , \pm 4\pi/Z, \cdots$ for a single-vortex. The BdG matrix in Eq.~(1) is numerically solved by spatial discretization with respect to $r$-axis, whose step is $0.01d$. The self-consistent gap equation for $\Delta({\bf r}) = e^{-i\theta}\Delta(r)$ is 
\begin{eqnarray}
\Delta(r) = \tilde{g}(r)\sum_{|E_{\bf q}| \le E_c} u_{\bf q}(r) v^{\ast}_{\bf q}(r) f(E_{\bf q}),
\label{eq:gap}
\end{eqnarray} 
where the effective coupling constant $\tilde{g}(r)$ is introduced in order to remove the ultra-violet divergence \cite{bruun,Bulgac}. Since we use the expression valid even for broken particle-hole symmetry, the sum in Eq.~(2) is done for all the eigenstates with both positive and negative eigenenergies up to the energy cutoff $\pm E_c = 200 \omega $ \cite{mizushima1}. In the BdG equation (\ref{eq:bdg}), the chemical potential is adjusted to fix the total particle number $N = N_{\uparrow} + N_{\downarrow} = 3000$. Here, we shall present the self-consistent results for the bare coupling constant $g/\omega d^3= -1.1, -1.2, -1.3, -1.4, -1.5$, corresponding to $-1.4 \le (k_Fa)^{-1} \le -0.7$ with the Fermi wave number $k_F$. These are also expressed as $\Delta _{0}(0)/E_F(0) = 0.15, 0.20, 0.23, 0.27, 0.32$ in the vortex-free state, respectively. The gap $\Delta_{0}(0)$ ($E_F(0)$) is $\Delta(r)$ (the Fermi energy $E_F$) at the trap center without a vortex in the population balance case.
The polarization $P$ is defined by $P=(N_{\uparrow}-N_{\downarrow})/(N_{\uparrow}+N_{\downarrow})$.
In this paper we only consider the $T=0$ case and neglect the external rotation energy.

We show the spatial profiles of the order parameter $\Delta(r)$ in Fig.~1. The vortex structure for the population imbalance corresponds to curve (1). It is seen that  the vortex core size is widen compared with that for the balance case (2). Towards the outer region $\Delta (r)$ oscillates and changes its sign several times to accommodate the excess majority component. This FFLO-like state is always stabilized irrespective of the values of  finite $P$ at $T=0$ as shown previously \cite{machida06}. Therefore for $P\neq 0$ at $T=0$ in the vortex solution $\Delta(r)$ changes its sign and oscillates.  As shown in the inset, which displays $|\Delta(r)|$ for curve (1) in a logarithmic scale to examine the detailed variation of the outer region, the oscillation pattern continues rather regularly to smaller scales of $\Delta$. This means that superfluidity is kept coherently throughout the whole system, even at the system edge where $|\Delta|$ is extremely small and the minority component is almost absent (See Fig.~2). Therefore the present solution for the imbalance case is quite different from a simple BCS-normal phase-separation in the local density approximation. 
This is consistent with that in our previous solution
for the vortex free case \cite{machida06} and with the work by Kinnunen {\it et al.} \cite{kinnunen}.

\begin{figure}[h]
\includegraphics[width=0.8\linewidth]{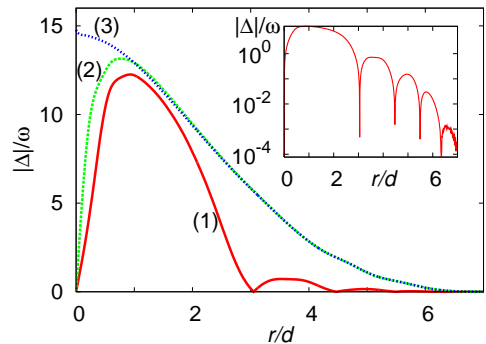}
\caption{(Color online) Order parameter profiles $|\Delta(r)|$ as a function of $r/d$ for $g/\omega d^3=-1.5$. (1) Vortex with population imbalance case $P=0.3$. (2) Vortex with population balance case $P=0$. (3) Vortex free with population balance case $P=0$. Logarithmic plot of (1) is shown in inset to emphasize the finer scale of $|\Delta(r)|$ in the outer region.}
\end{figure}

In Fig.~2 we show the density profiles; the total $n(r)$, the majority $n _{\uparrow}(r)$ and minority $n _{\downarrow}(r)$ components, and the magnetization $m(r) =n _{\uparrow}(r) - n _{\downarrow}(r)$ for the solution corresponding to curve (1) in Fig.~1. It is seen that  the majority component exclusively occupies the central region while the minority component remains depleted. It should be noted that the quantum depletion of the total number $n(r)$ at $T=0$ is substantially smaller and incomplete compared with the balance case as seen shortly in Fig.~4.  The magnetization shows a peak at the vortex center. In the intermediate region $0.5 \le r/d \le 2.5$ the magnetization is almost vanishing, resulting in the enhancement of $n _{\downarrow}(r)$. This leads to the ``bimodal'' structure of $n _{\downarrow}(r)$, which is observed experimentally
in the vortex free cases \cite{mit1,mit2}.
This bimodal structure is confirmed  by our previous calculation in the vortex free case \cite{machida06}.
In the outer region $r \ge 5d$ in Fig.~2 there exists only the majority component $n _{\uparrow}(r)$ and the magnetization is fully polarized there.

\begin{figure}[h]
\includegraphics[width=0.8\linewidth]{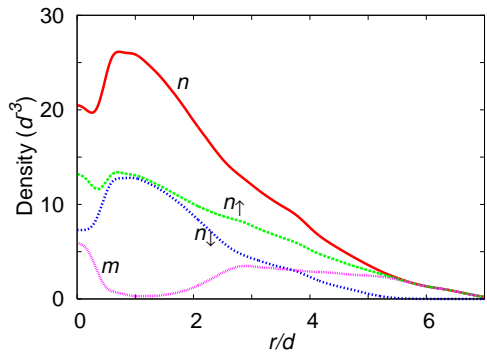}
\caption{(Color online) Density and polarization profiles as a function of $r/d$ corresponding to curve (1) ($P=0.3, g/\omega d^3 =-1.5$) in Fig.~1. $n(r)=n_\uparrow(r)+n_\downarrow(r)$: total density, $n_\uparrow (r) (n_\downarrow(r))$: up (down) spin density and $m(r)$: magnetization.}
\end{figure}

In Fig.~3 we show density plots for $n _{\uparrow }(r), n_{\downarrow}(r)$ and $n(r)$ of Fig.~2 where the vortex core is located at the center. The contrast in $n _{\downarrow}(r)$ and $n(r)$ is clearer than that of $n _{\uparrow}(r)$, which is dim. Namely in the imbalance vortex, the minority (majority) component is visible as in Fig.~3(b) (invisible as in Fig.~3(a)).
We also show the density plot of $n(r)$ for the balance case 
for comparison in Fig.~3(d). It is seen that the contrast of the balance case is clearer compared with that for the 
imbalance case in Fig.~3(c).
This is because the quantum depletion of $n(r)$ occurs at the core both for $n _{\uparrow}(r)$ and $n _{\downarrow}(r)$.  These results in Fig.~3 are checked experimentally by direct ``in situ'' method \cite{mit2,mit3} without resorting to BEC mapping from the BCS side, combined with phase contrast imaging \cite{mit3}. When the BEC mapping is used \cite{mit1}, the density contrast of vortices may be modified.
We note in the passing that the BEC case $n(r) =0$ at the vortex center, because $n(r)$ is the order parameter and at the vortex center $n(0)$ must vanish because of the presence of the phase singularity due to phase winding.

\begin{figure}[h]
\includegraphics[width=0.8\linewidth]{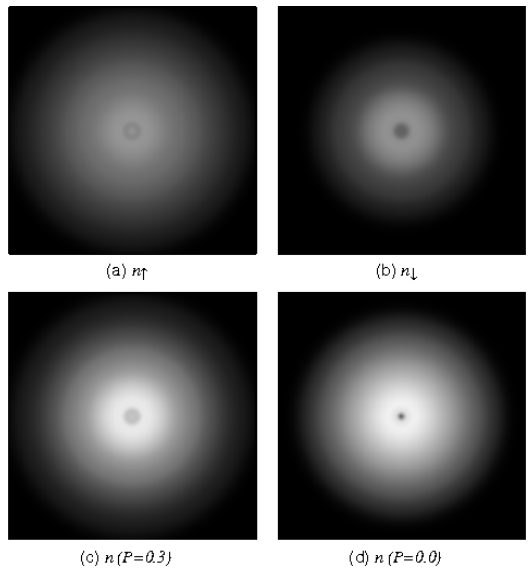}
\caption{Density maps of (a) $n_\uparrow(r)$, (b) $n_\downarrow(r)$ and (c) $n(r)$, corresponding to Fig.~2. (d) $n(r)$ for $P=0$ and $g/\omega d^3=-1.5$. By comparing the imbalance case (c) and the balance case (d), the core in (d) is clearly seen as a small but sharp dot at the center. The field of view is $14.0 \times 14.0$ in units of $d$.}
\end{figure}

The polarization $P$ dependence of the total density profiles is displayed in Fig.~4. In the balance case ($P=0$) substantial quantum depletion at the vortex core is seen. As $P$ increases, the core density is progressively filled out and the vortex filling becomes complete, thus the vortex tends to be ``invisible'' in the density contrast experiment. The core radius defined at the density maximum point from the trap center increases with increasing $P$, then the ``core size'' expands. 

\begin{figure}[h]
\includegraphics[width=0.8\linewidth]{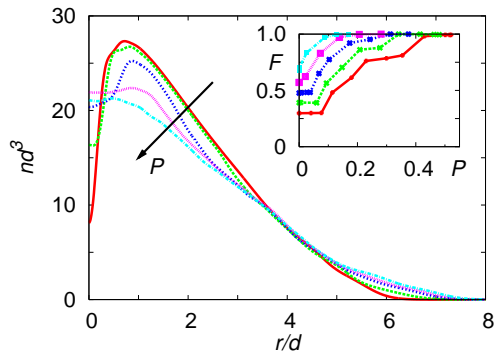}
\caption{(Color online) Density profiles of the total number $n(r)$ for various polarizations: $P=0,0.11,0.35,0.43,0.52$ from top to bottom. The inset shows the filling factor $F=n(0)/n_{max}$ defined by the ratio between the maximum value $n_{max}$ and $n(r=0)$ for various couplings ($g/\omega d^3=-1.5,-1.4,-1.3,-1.2,-1.1$ from bottom to top).}
\end{figure}

The inset of Fig.~4 shows the $P$-dependence of the core filling factor $F=n(0)/n_{max}$ defined by the density ratio between the density $n(0)$ at the core center and the maximum density $n_{max}$. It demonstrates how visibility becomes lower with $P$. Due to the quantum depletion at the core, which is proportional to $\Delta_0(0)/E_F$ \cite{hayashi}, the filling factor $F$  becomes smaller as the strength of the attractive interaction $|g|$ increases, i.e. approaching the Feshbach resonance point.  The density contrast quickly diminishes with $P$, implying that vortex becomes invisible with $P$ for general $g$. As seen from the inset of Fig.~4 the filling factor $F$ as a function of $P$ changes in a step-like manner, particularly for larger $|g|$.
This comes from the fact that the vortex bound states are discretized. Thus as $P$ varies, the occupation of these quantized states is progressively changed, leading
to the step-like variation of the occupation number.

These results might be relevant for the interpretation of the vortex experiment in the imbalance case, especially, around the outer region. In the actual experiment the vortex visibility is influenced by the formation of FFLO as follows:
(i) In the outer region, FFLO pairing survives with small amplitude and
(ii) it is also always accompanied with a large amount of polarization.
By these effects, vortices around outer region in the FFLO state can be invisible.
The vortices at the center (outer) correspond to the present single vortex study for smaller (larger) $P$. Hence, we expect general tendency that in the imbalance case the vortices in the center (outer) region are visible (invisible).
 
\begin{figure}[h]
\includegraphics[width=0.75\linewidth]{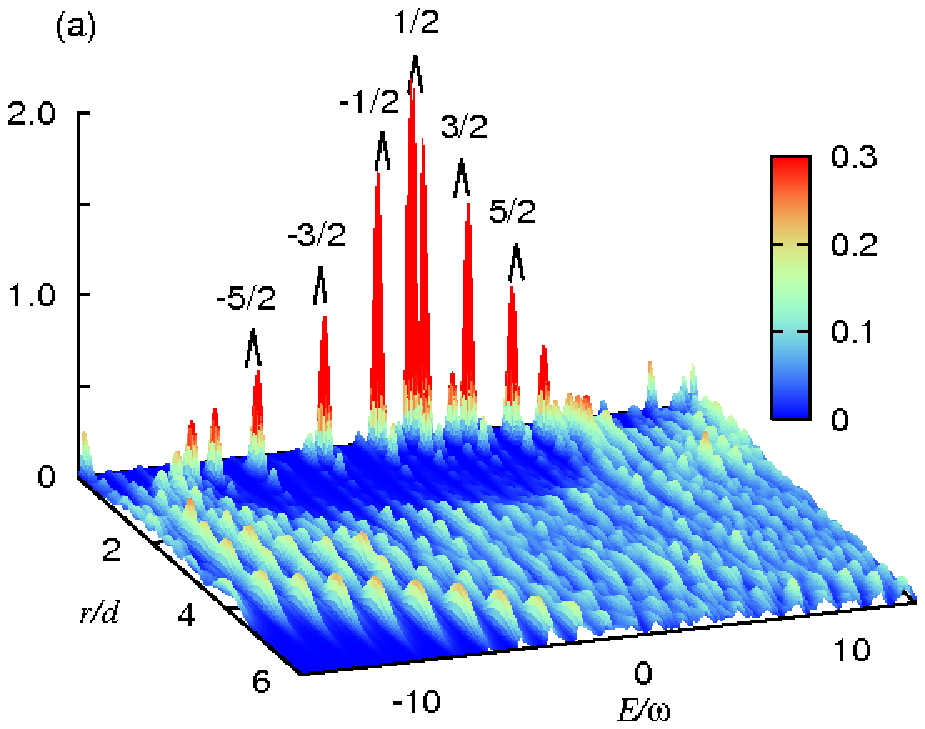}
\vspace*{3mm}\ 
\includegraphics[width=0.7\linewidth]{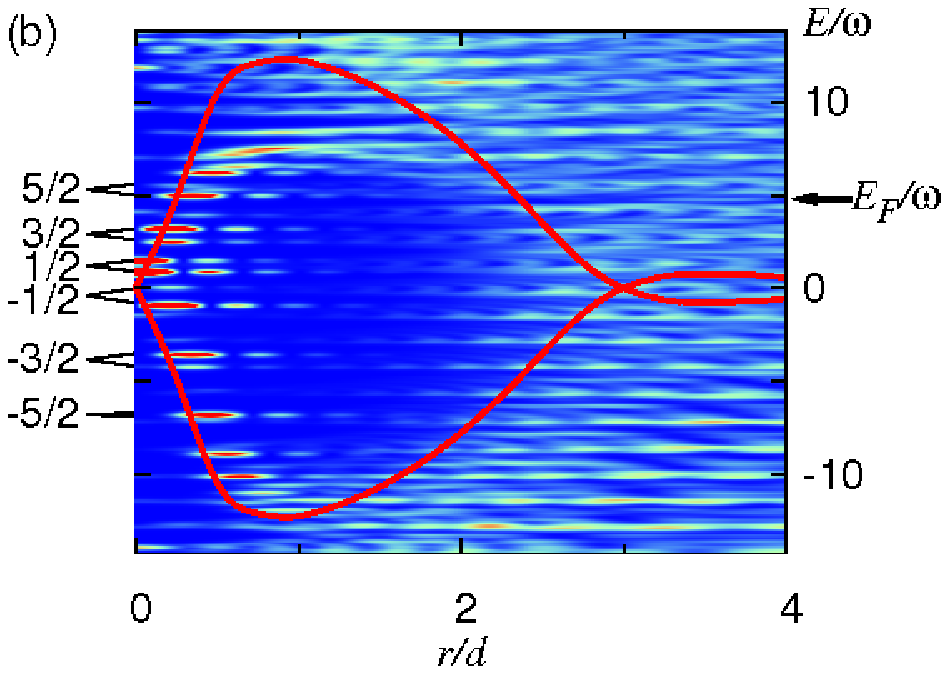}
\caption{(Color online) Spectral evolutions for the majority species (up-spin atoms) $N_\uparrow(r,E)$ with $q_z=0$. Stereographic view (a) and density map (b). $P=0.3$ and $g/\omega d^3=-1.5$. The labels show the angular momentum $l$ for each CdGM bound state. In (b) the order parameter $\pm \Delta(r)$ is overlaid. Energy is normalized by the harmonic frequency $\omega$. $E_F$ is the Fermi level position.}
\end{figure}

In order to understand why the total number $n(r)$, in particular the majority species $n_{\uparrow}(r)$ fills out the core region, we examine the local density of states with $q_{z} =0$, namely, $N_{\uparrow}(r,E) = \sum _{\bf q} |u_{\bf q}(r)|^2 \delta(E-E_{\bf q})$ and $N_{\downarrow}(r,E) = \sum _{\bf q} |v_{\bf q}(r)|^2 \delta(E+E_{\bf q})$. The spectral evolution of $N_{\uparrow}(r,E)$ for the majority species near the Fermi level $E_F$ and the corresponding density plot are displayed respectively in Figs.~5(a) and 5(b). It is seen clearly from Fig.~5(a) that near the vortex core $r=0$ and at the lower energy region around $E=0$, a series of the core CdGM bound states \cite{caroli}, i.e. more broadly the Andreev bound states, are located, whose angular momentum is characterized by half-integers $\ell = \pm \frac{1}{2}, \pm\frac{3}{2}, \cdots$. At the core center $r=0$ only the CdGM bound state labeled by $\ell = \frac{1}{2}$ with a positive energy has spectral amplitude and the others have no spectral weight at $r=0 $ \cite{hayashi}. This bound state is unoccupied when $P=0$. Since for $P\neq 0$ it is occupied because of the upward Fermi level shift, the majority species $n_{\uparrow}(r=0)$ becomes able to fill in the core. For the minority species this state is unoccupied, thus $n_{\downarrow}(r)$ remains suppressed at the core. Since the high angular momentum states with $l=3/2, 5/2, \cdots$ are  progressively occupied for the majority species, $n_{\uparrow}(r)$ further increases near the core. Beyond a certain $P$ value, which depends on the coupling constant, the depletion of the core becomes absent 
and the core filling is complete (see inset of Fig.~4).

We should notice from Fig.~5(a) that each CdGM bound state is split and has the double peak structure. This can be understood as follows: We show the spectral density map of $N_{\uparrow}(r,E)$ together with the order parameter profile $\pm \Delta(r)$ in Fig.~5(b). At the node ($r/d = 3.0$) where $\Delta(r)=0$, there appear the zero-energy bound states, i.e. the so-called midgap state associated with FFLO. This midgap state accompanies several excited states around the node because quasi-particles feel a local confining potential created by the nodes of $|\Delta(r)|$ \cite{baranov}. Although these Andreev bound states are localized around the nodes, they can resonate with the CdGM state with $\ell = \frac{1}{2}$ bounded by the  vortex if these energies match each other. Other CdGM states with higher angular momentum may also resonate with the surface harmonic states which are originally localized at the surface. These new ``split'' bound states further enhance the $n_{\uparrow}$ occupation and the magnetization around the vortex core.  Physically this enhancement is reasonable
because these newly created resonant bound states are advantageous in view of the potential energy gain at the
center.

In summary, we have shown by solving microscopic BdG equation self-consistently that the vortex core 
in Fermion superfluids tends to
be filled in by the majority species when the population of the two species is imbalance. It makes the vortex core image unclear via the density profile measurement. 
The vortex core can still be visible through the selective measurement of the minority density, using phase contrast imaging.
We have given its physical reasons in terms of the Caroli-de Gennes-Matricon bound states. These predictions can be checked experimentally by direct ``in situ'' method \cite{mit2,mit3} without resorting to BEC mapping \cite{mit1} from the BCS side, combined with newly developed phase contrast imaging \cite{mit3}.

We would like to thank  M.W. Zwierlein and N. Hayashi for stimulating discussions.

\end{document}